\documentstyle[12pt]{article}

\catcode`\@=11
\long\def\@makefntext#1{ 
\protect\noindent \hbox to 3.2pt {\hskip-.9pt  
$^{{\ninerm\@thefnmark}}$\hfil}#1\hfill} 

\def\thefootnote{\fnsymbol{footnote}}
 \def\@makefnmark{\hbox to 0pt{$^{\@thefnmark}$\hss}}  
	
\def\ps@myheadings{\let\@mkboth\@gobbletwo
\def\@oddhead{\hbox{} 
\rightmark\hfil\ninerm\thepage}   
\def\@oddfoot{}\def\@evenhead{\ninerm\thepage\hfil 
\leftmark\hbox{}}\def\@evenfoot{}
\def\sectionmark##1{}\def\subsectionmark##1{}}
\def\etal{et al.}
\def\hmpc{\rm \; h^{-1}\; Mpc}

\begin{document}

\newcommand{\symbolfootnote}{\renewcommand{\thefootnote}
	{\fnsymbol{footnote}}}
\renewcommand{\thefootnote}{\fnsymbol{footnote}}
\newcommand{\alphfootnote}
	{\setcounter{footnote}{0}
	 \renewcommand{\thefootnote}{\sevenrm\alph{footnote}}}

\newcounter{sectionc}\newcounter{subsectionc}\newcounter{subsubsectionc}
\renewcommand{\section}[1] {\vspace{0.6cm}\addtocounter{sectionc}{1} 
\setcounter{subsectionc}{0}\setcounter{subsubsectionc}{0}\noindent 
	{\bf\thesectionc. #1}\par\vspace{0.4cm}}
\renewcommand{\subsection}[1] {\vspace{0.6cm}\addtocounter{subsectionc}{1} 
	\setcounter{subsubsectionc}{0}\noindent 
	{\it\thesectionc.\thesubsectionc. #1}\par\vspace{0.4cm}}
\renewcommand{\subsubsection}[1] {\vspace{0.6cm}\addtocounter{subsubsectionc}{1}
	\noindent {\rm\thesectionc.\thesubsectionc.\thesubsubsectionc. 
	#1}\par\vspace{0.4cm}}
\newcommand{\nonumsection}[1] {\vspace{0.6cm}\noindent{\bf #1}
	\par\vspace{0.4cm}}
					         
\newcounter{appendixc}
\newcounter{subappendixc}[appendixc]
\newcounter{subsubappendixc}[subappendixc]
\renewcommand{\thesubappendixc}{\Alph{appendixc}.\arabic{subappendixc}}
\renewcommand{\thesubsubappendixc}
	{\Alph{appendixc}.\arabic{subappendixc}.\arabic{subsubappendixc}}

\renewcommand{\appendix}[1] {\vspace{0.6cm}
        \refstepcounter{appendixc}
        \setcounter{figure}{0}
        \setcounter{table}{0}
        \setcounter{equation}{0}
        \renewcommand{\thefigure}{\Alph{appendixc}.\arabic{figure}}
        \renewcommand{\thetable}{\Alph{appendixc}.\arabic{table}}
        \renewcommand{\theappendixc}{\Alph{appendixc}}
        \renewcommand{\theequation}{\Alph{appendixc}.\arabic{equation}}
        \noindent{\bf Appendix \theappendixc #1}\par\vspace{0.4cm}}
\newcommand{\subappendix}[1] {\vspace{0.6cm}
        \refstepcounter{subappendixc}
        \noindent{\bf Appendix \thesubappendixc. #1}\par\vspace{0.4cm}}
\newcommand{\subsubappendix}[1] {\vspace{0.6cm}
        \refstepcounter{subsubappendixc}
        \noindent{\it Appendix \thesubsubappendixc. #1}
	\par\vspace{0.4cm}}

\def\abstracts#1{{
	\centering{\begin{minipage}{30pc}\tenrm\baselineskip=12pt\noindent
	\centerline{\tenrm ABSTRACT}\vspace{0.3cm}
	\parindent=0pt #1
	\end{minipage} }\par}} 

\newcommand{\bibit}{\it}
\newcommand{\bibbf}{\bf}
\renewenvironment{thebibliography}[1]
	{\begin{list}{\arabic{enumi}.}
	{\usecounter{enumi}\setlength{\parsep}{0pt}
\setlength{\leftmargin 1.25cm}{\rightmargin 0pt}
	 \setlength{\itemsep}{0pt} \settowidth
	{\labelwidth}{#1.}\sloppy}}{\end{list}}

\topsep=0in\parsep=0in\itemsep=0in
\parindent=1.5pc

\newcounter{itemlistc}
\newcounter{romanlistc}
\newcounter{alphlistc}
\newcounter{arabiclistc}
\newenvironment{itemlist}
    	{\setcounter{itemlistc}{0}
	 \begin{list}{$\bullet$}
	{\usecounter{itemlistc}
	 \setlength{\parsep}{0pt}
	 \setlength{\itemsep}{0pt}}}{\end{list}}

\newenvironment{romanlist}
	{\setcounter{romanlistc}{0}
	 \begin{list}{$($\roman{romanlistc}$)$}
	{\usecounter{romanlistc}
	 \setlength{\parsep}{0pt}
	 \setlength{\itemsep}{0pt}}}{\end{list}}

\newenvironment{alphlist}
	{\setcounter{alphlistc}{0}
	 \begin{list}{$($\alph{alphlistc}$)$}
	{\usecounter{alphlistc}
	 \setlength{\parsep}{0pt}
	 \setlength{\itemsep}{0pt}}}{\end{list}}

\newenvironment{arabiclist}
	{\setcounter{arabiclistc}{0}
	 \begin{list}{\arabic{arabiclistc}}
	{\usecounter{arabiclistc}
	 \setlength{\parsep}{0pt}
	 \setlength{\itemsep}{0pt}}}{\end{list}}

\newcommand{\fcaption}[1]{
        \refstepcounter{figure}
        \setbox\@tempboxa = \hbox{\tenrm Fig.~\thefigure. #1}
        \ifdim \wd\@tempboxa > 6in
           {\begin{center}
        \parbox{6in}{\tenrm\baselineskip=12pt Fig.~\thefigure. #1 }
            \end{center}}
        \else
             {\begin{center}
             {\tenrm Fig.~\thefigure. #1}
              \end{center}}
        \fi}

\newcommand{\tcaption}[1]{
        \refstepcounter{table}
        \setbox\@tempboxa = \hbox{\tenrm Table~\thetable. #1}
        \ifdim \wd\@tempboxa > 6in
           {\begin{center}
        \parbox{6in}{\tenrm\baselineskip=12pt Table~\thetable. #1 }
            \end{center}}
        \else
             {\begin{center}
             {\tenrm Table~\thetable. #1}
              \end{center}}
        \fi}

\def\@citex[#1]#2{\if@filesw\immediate\write\@auxout
	{\string\citation{#2}}\fi
\def\@citea{}\@cite{\@for\@citeb:=#2\do
	{\@citea\def\@citea{,}\@ifundefined
	{b@\@citeb}{{\bf ?}\@warning
	{Citation `\@citeb' on page \thepage \space undefined}}
	{\csname b@\@citeb\endcsname}}}{#1}}

\newif\if@cghi
\def\cite{\@cghitrue\@ifnextchar [{\@tempswatrue
	\@citex}{\@tempswafalse\@citex[]}}
\def\citelow{\@cghifalse\@ifnextchar [{\@tempswatrue
	\@citex}{\@tempswafalse\@citex[]}}
\def\@cite#1#2{{$\null^{#1}$\if@tempswa\typeout
	{IJCGA warning: optional citation argument 
	ignored: `#2'} \fi}}
\newcommand{\citeup}{\cite}

\def\fnm#1{$^{\mbox{\scriptsize #1}}$}
\def\fnt#1#2{\footnotetext{\kern-.3em
	{$^{\mbox{\sevenrm #1}}$}{#2}}}

\font\twelvebf=cmbx10 scaled\magstep 1
\font\twelverm=cmr10 scaled\magstep 1
\font\twelveit=cmti10 scaled\magstep 1
\font\elevenbfit=cmbxti10 scaled\magstephalf
\font\elevenbf=cmbx10 scaled\magstephalf
\font\elevenrm=cmr10 scaled\magstephalf
\font\elevenit=cmti10 scaled\magstephalf
\font\bfit=cmbxti10
\font\tenbf=cmbx10
\font\tenrm=cmr10
\font\tenit=cmti10
\font\ninebf=cmbx9
\font\ninerm=cmr9
\font\nineit=cmti9
\font\eightbf=cmbx8
\font\eightrm=cmr8
\font\eightit=cmti8


\centerline{\tenbf THE ESO ALL--SOUTHERN--SKY REDSHIFT SURVEY}
\baselineskip=16pt
\centerline{\tenbf OF ROSAT CLUSTERS OF GALAXIES\footnote{to appear
in {\tenit Wide--Field Spectroscopy and the Distant
Universe,} S.J. Maddox and A. Arag\'on-Salamanca eds., World Scientific,
Singapore.}}
\vspace{0.8cm}
\baselineskip=12pt

\noindent{\tenrm L. Guzzo$^{(a)}$\footnote{e-mail: guzzo@astmim.mi.astro.it}, 
H. B\"ohringer$^{(b)}$, U. Briel$^{(b)}$, 
G. Chincarini$^{(a,c)}$, 
C.A. Collins$^{(d)}$, S. De~Grandi$^{(c)}$, H. Ebeling$^{(e)}$, 
A.C. Edge$^{(e)}$, 
D. Neumann$^{(b)}$, S. Schindler$^{(f)}$, P. Schuecker$^{(g)}$, 
W. Seitter$^{(g)}$, 
P. Shaver$^{(h)}$, P. Vettolani$^{(i)}$, R. Cruddace$^{(j)}$, 
A.C. Fabian$^{(e)}$,
H. Gursky$^{(j)}$, R. Gruber$^{(b)}$, G. Hartner$^{(b)}$, 
H.T. MacGillivray$^{(k)}$, 
D. Maccagni$^{(l)}$, M. Pierre$^{(m)}$, K. Romer$^{(n)}$, W. Voges$^{(b)}$, 
J. Wallin$^{(j)}$, A. Wolter$^{(a)}$, G. Zamorani$^{(o)}$}

\vspace{0.4cm}
\leftline{\tenit $^{(a)}$Osservatorio Astronomico di Brera, Milano/Merate, Italy}
\leftline{\tenit $^{(b)}$Max-Planck-Institut f\"ur Extraterrestrische Physik, Garching, Germany}
\leftline{\tenit $^{(c)}$Dipartimento di Fisica, Universit\`a di Milano, Milano, Italy}
\leftline{\tenit $^{(d)}$Liverpool John-Moores University, Liverpool, U.K.}
\leftline{\tenit $^{(e)}$Institute of Astronomy, Cambridge, UK}
\leftline{\tenit $^{(f)}$Max-Planck-Institut f\"ur Astrophysik, Garching, Germany}
\leftline{\tenit $^{(g)}$Astronomisches Institut der Universit\"at M\"unster, M\"unster, Germany}
\leftline{\tenit $^{(h)}$European Southern Observatory, Garching, Germany}
\leftline{\tenit $^{(i)}$Istituto di Radioastronomia del CNR, Bologna, Italy}
\leftline{\tenit $^{(j)}$Naval Research Lab., Washington D.C., USA}
\leftline{\tenit $^{(k)}$Royal Observatory Edinburgh, Edinburgh, U.K.}
\leftline{\tenit $^{(l)}$Istituto di Fisica Cosmica del CNR, Milano, Italy}
\leftline{\tenit $^{(m)}$CEN Saclay, Gif-sur-Yvette, France}
\leftline{\tenit $^{(n)}$Northwestern University, Evanston, Il, U.S.A.}
\leftline{\tenit $^{(o)}$Osservatorio Astronomico, Bologna, Italy}

\vspace{0.9cm}

\abstracts{We discuss the rationale and the present status of a large 
redshift survey aiming at measuring distances for all the clusters 
of galaxies detected by the ROSAT X--ray All--Sky Survey in the 
southern hemisphere, with flux 
larger than $\sim 2\times 10^{-12}$ erg s$^{-1}$ cm$^{-2}$.  
The survey is being performed using 
three ESO telescopes in parallel.   The final sample will contain about 700 
clusters.  It will represent a unique database for the study of the 
spatial distribution of clusters, their X--ray luminosity function 
and their intrinsic physical properties.   
Serendipitous results obtained so far include two new 
gravitational arc systems in two clusters of galaxies.  These findings, 
obtained on rather short exposures, confirm the advantage of searching 
for arcs in X--ray selected clusters.
}

\newpage
\vfil
\rm\baselineskip=14pt
\section{Clusters as Cosmological Probes}

In the hierarchy of cosmological structures, clusters of galaxies are the
largest well--defined building blocks and represent the deepest gravitational
potential wells that can be found in the Universe.   By well--defined, one
means that they have clearly separated from the Hubble expansion and
recollapsed to a nearly (or fully, in some cases) virialized object.
In addition to the galaxy content, these ``cosmological sinks" are filled 
with large masses of gas, heated at temperatures of 1--10 keV and emitting
luminosities in the soft X--ray band around $10^{44}$ erg s$^{-1}$.  
As a consequence, X--ray
emission from clusters can be detected up to very large distances (see
e.g. ref. 1).  This specific property makes clusters of galaxies 
very interesting as tracers of the large--scale structure of the Universe,
both statically (to characterize present--time structure on very large
scales), and dynamically (to study the evolution of clustering).

As large--scale structure tracers, clusters of galaxies present a number
of advantages.   The first is that they basically show a nearly frozen --
though thresholded at high contrast -- picture of the initial fluctuation
field.  Indeed, with typical mean separations $\sim 10 \hmpc$, they 
map scales which are essentially still in their linear regime of growth.
Any clustering statistics derived from the distribution of clusters is
therefore in principle easier to be compared to the model predictions.
Observationally, clusters
are ``cheap" tracers, in the sense that for covering a given volume of
space they require a much smaller number of galaxy redshifts than when
using single galaxies.  There is about a factor of 100 between the number of
redshifts required by the two approaches, depending on the number of galaxies
used for obtaining the mean cluster velocity.   While cluster redshift
surveys have been able to explore three--dimensional volumes already covering 
typical scales around $\sim 300 \hmpc$, similar volumes will be reached
only by next generation single galaxy surveys, thanks to the advent of
super--multiplexing spectrographs and/or dedicated telescopes (see the 
contributions by Gunn and by Taylor in this same volume).

The primary reference list for studies of galaxy clusters during the
last few decades, has been the catalogue compiled by Abell in 1954 
(together with its more recent extension to the South, the ACO catalogue).  
Classic results on the large--scale distribution of clusters were
obtained on richness--limited subsamples of this compilation,
like e.g. the estimate of the cluster--cluster correlation function
$\xi_{cc}(r)$\cite{bs83,klko}.
However, the robustness of these results was often questioned 
in relation to the intrinsically subjective nature of the 
visual cluster selection, and also to possible 
projection effects artificially boosting the number of angularly close pairs
when selecting richness--limited samples\cite{luc,sut,dek}.
The construction of automatic cluster catalogues 
from digitized galaxy data, like the Edinburgh/Durham Cluster Catalogue 
(EDCC)\cite{lmsd}, represented a first step towards obtaining a 
statistically complete and unbiased sample of clusters.  In fact, 
clustering analyses from the EDCC\cite{bob92}, and from the similar APM 
cluster catalogue\cite{dalt} showed a correlation function more 
isotropic and with a smaller amplitude than those estimated from Abell
samples.

Although the use of specific algorithms considerably reduced projection
effects in the automatic catalogues, the selection of the clusters 
from the projected two--dimensional distribution of galaxies inevitably 
carries a percentage of spurious rich cluster produced by the chance
superposition of different structures along the line of sight ($\sim 
10$\% for the EDCC\cite{col94}).  The powerful X--ray emission
from galaxy clusters provides a much better way to intrinsically define 
a cluster and construct virtually projection--free catalogues.   
This is also helped by the fact that the X-ray emission produced by thermal
bremsstrahlung is a quadratic function of the density. If the
emitting gas follows the same radial profile as the galaxy population,
it is clear that the X--ray emission will be roughly proportional to
the richness squared, providing a more compact system in
the X--ray band rather than in the optical. (In fact, the proportionality
law between luminosity and density could even be a power higher than two
if the cluster contains a cooling flow in its center).

Another related issue on which a large sample of clusters selected in 
X--rays would be invaluable is that of the `local' X--ray luminosity function 
of clusters.
This is directly related to the distribution function of the mass in the 
Universe, and thus it is just another route to probe the power spectrum of 
density fluctuations on very large scales (e.g. ref.~11).
In addition, a reliable determination of the present--epoch X--ray luminosity
function is also crucial to the study of the evolution of the function itself.
The results obtained from the EXOSAT and EINSTEIN data point towards
the existence of rapid evolution\cite{giolf,edglf}.  While observations of
new homogeneous samples of distant clusters are needed to test for this effect, 
a large, unbiased local sample is also essential for proper comparison.

The dynamical state of clusters both at the present epoch and as a function
of cosmic time (also intimately related to the evolution of the X--ray 
luminosity function), is a further important observation for constraining 
cosmological models.  The mean degree of virialization at the present epoch, 
that can be studied through the correlation of X--ray luminosity and galaxy 
velocity dispersion, or by comparing the X--ray and optical morphologies, 
is clearly dependent on the cosmological scenario.  Again, significant 
results on this topic depend on the size and quality of the available
X--ray cluster sample.

It becomes therefore clear how an all--sky survey in the X--ray 
band, like that performed by the ROSAT satellite, represents a unique 
opportunity for constructing a large, homogeneous and statistically 
complete catalogue of galaxy clusters to address the above discussed
issues. 

\section{The ROSAT All--Sky Survey}

Between August 1990 and January 1991, the ROSAT satellite performed
the first ever all--sky survey using an imaging X--ray telescope 
\cite{tru}.   In the configuration used for the survey, 
the ROSAT telescope has a field of view of 2 degrees, covering
the energy band between 0.1 and 2.4 keV, with a resolution of 20 arcsec on
axis which degrades to 2 arcmin at the field edges.   The mean exposure
time was around 600 sec, with peaks of $\sim 40000$ sec at the
ecliptic poles\cite{vog,bhr94}.
Further details on the instrument and on the ROSAT 
All--Sky Survey (RASS) are presented in these proceedings by Hasinger.

The first processing of the whole survey with the so--called Standard
Analysis Software System (SASS) produced a database of some 50000 sources.
The sensitivity limit can be placed around $10^{-12}$ erg s$^{-1}$ cm$^{-2}$,
which, using results from previous surveys\cite{evr}, gives a number of 
4000--5000 clusters of galaxies expected 
in the whole survey\cite{bhr91,bhr94}.
Most of these clusters are expected to be at a redshift comprised 
between 0.1 and 0.2, with a few very luminous objects at $z\ge0.3$.   

Optical follow--up of ROSAT clusters started early after completion
of the SASS on subsets of the all--sky sample.  An overview of these
projects has been given by B\"ohringer\cite{bhr94}.  In particular, 
an extensive redshift survey has been
performed in the region of the South Galactic Pole (SGP), 
producing an estimate of the correlation function of X--ray clusters\cite{rmr}
which is in good agreement with the above mentioned results from the 
automatic catalogues (see also the contribution by Collins in this volume).

However, what makes the ROSAT survey unique is its all--sky distribution,
which is ideal for accurate studies of large--scale structure, opening 
the way to clustering analyses on scales of the order of 1000 $\hmpc$.  

\section{An All--Southern Sky Redshift Survey of X--Ray Revealed Clusters}

To fully exploit the above described potential of the RASS, in 1991 we
proposed to ESO a key programme to survey spectroscopically all the ROSAT
clusters down to a certain flux limit over the entire southern sky ($|b|\ge 
20^{\circ}$).  The project was approved, awarding 90 nights equally 
distributed among the 3.6~m, 2.2~m and 1.5~m telescopes.  

\vspace{8cm}
\fcaption{The sky distribution of the cluster candidates above a SASS
count rate of 0.1 cts s$^{-1}$.  Filled circles (248) mark the objects
for which redshifts have been already observed in the program discussed
here or in the Hydra and SGP projects.   Dark grey circles (132) 
are clusters with redshift from the literature, while light grey circles
are still to be observed.}

\medskip
Our observational 
goal is to obtain a reliable mean redshift for all clusters down to a SASS
count rate limit of 0.08 cts s$^{-1}$, corresponding to a flux limit
of 1.6--2$\times 10^{-12}$ erg s$^{-1}$ cm$^{-2}$.  To this flux, the present
cluster candidate list -- obtained from the combined 
cluster identification procedures described below -- contains
$\sim~700$ objects.   Nearly 300 of these clusters have already a 
measured redshift, either from the literature or from one of the
early redshift follow--ups in restricted regions (like, e.g., the
SGP project).   Fig.~1 shows the sky distribution of the candidates
above a SASS count rate of 0.1 cts s$^{-1}$.  

\subsection{Cluster Identification}

A complete
optical identification of all the RASS sources is presently performed only 
within a few selected regions, by a separate project involving 
the use of a large amount of telescope time.   In the case of our
survey, the cluster identification from the X--ray source 
catalogue produced by the SASS has to follow a different approach.  
The aim is to construct a sample which is complete for X--ray clusters, 
i.e. that contains virtually
all the RASS sources associated with a galaxy cluster.  
The basic idea is to merge together the products of a few
different criteria, using relatively broad boundaries to avoid 
discarding genuine X--ray clusters.  The price to be paid is that 
of including in the candidate list also a number of spurious objects
that will have to be discarded through a direct screening 
of the ESO/SRC sky survey plates.   This means that we are requesting 
that some signature of the cluster is present on the optical sky survey 
images.  This introduces in practice a redshift cut around $z\simeq0.3$.
This is not problematic for the completeness of our sample, given its 
flux limit and the consequently expected redshift distribution which 
peaks at $z$'s between 0.1 and 0.2.

The main preidentification method adopted for the southern hemisphere, 
is based on the comparison of the RASS source list
with an optical galaxy catalogue ($b_j \le 20.5$), constructed in Edinburgh 
with the COSMOS machine digitizing the whole southern sky ESO/SRC J survey 
plates.
All the SASS sources are correlated with the galaxy positions in this
catalogue, and also with a list of galaxy concentrations (i.e. `clusters'),
automatically
constructed from this optical database.  The number of galaxies around 
each X--ray source is compared to the mean number expected for random
fields, and overdensities above a given threshold are flagged.  The SASS
source list is also correlated with the Abell/ACO and Zwicky catalogues
of clusters.

Further potential cluster candidates are all the X--ray sources 
which are recognized as extended by the SASS: 70\% of these are found to be 
real clusters.  The rest are usually very bright stars or AGNs (for which 
small deviations between the theoretically assumed point spread function and 
the observed one lead to a spurious significance for the extent), or
nearby galaxies.  This is evidently the only selection criterion which
is solely based on the properties of the X--ray emission.  Note however
that at z=0.1 many clusters have already an angular core radius
$\sim 1$ arcmin, i.e. below the resolution of the instrument, and are
therefore detected as point sources.  It is interesting to note that 
a selection based only on the X--ray extension (though a very clear--cut
one), recovers only one--third of our present sample of ROSAT detected 
clusters.  

There is obviously an overlap between the sources found by the different
approaches.  In this sense, the different ways of identifying the clusters 
represent an important exercise to check the relative completeness and biases 
of the different techniques used.  We expect the present sample, 
defined within the limits discussed in the beginning of this section, 
and obtained through the screening and the merging of the different methods,
to be more than 80\% complete.  This figure,
which is rather uncertain at the moment, is possibly pessimistic and 
should inevitably improve with the refinements of the analysis of the RASS data.
While the redshift survey is progressing steadily, we are working
in parallel to different ways of detecting sources and estimating their 
fluxes from the RASS data.  A notable example is the so--called VTP 
method\cite{ebel}
(`Voronoi Tessellation and Percolation', see Ebeling, these proceedings).
An important step forward in this sense will definitely
be represented by the second--generation Standard  Analysis of the RASS data
(SASS II),
which is expected before the end of 1994 and should improve significantly the 
detection efficiency and flux estimate for extended sources.

As a final remark, it is important to remind that not everything is golden
also when selecting clusters in X--rays.  In particular, a main difference
with respect to optically selected samples is 
that once the X--ray source has been associated to a cluster
of galaxies, the possibility that part (or most) of its flux
comes from a single galaxy within the cluster (e.g. a BL Lac object or 
a Seyfert galaxy) has still to be taken into consideration.  To this end
(in addition to looking, when enough resolved, at the X--ray and optical 
morphologies), we devote special care to the presence of high--excitation 
emission lines in cluster galaxies as indicator of high--energy nuclear 
activity.

\subsection{Observational Goals and Strategy}

Our main observational goal is to measure reliable
cluster redshifts.  This implies observing spectroscopically at least 5
redshifts per cluster, possibly 10 or more.  A practical benefit of
this is that a velocity dispersion will also be obtained for many
of the observed clusters.   Another important point we already mentioned 
is the careful check for dubious identifications and possible
AGN contaminants.   These goals are met first of all by the strategy of using
multi--object spectroscopy (MOS) at the 3.6~m telescope equipped
with the high--throughput spectrograph EFOSC.  With this instrument,
around 10--20 redshifts per field are typically secured at once on
moderately distant and/or compact clusters, with around 10 clusters
observable per night.   On more nearby, looser systems, we use
single--slit spectroscopy (at the 2.2~m and 1.5~m telescopes), with,
respectively, an efficiency of $\sim 20$ and 
$\sim 10$ slit positions observed each night.

\newpage
\section{Present Status, Serendipitous Results and the Future}

So far (mid-1994), we have invested around 30\% of the total allocated
time.  A total of 158 cluster candidates have been observed
from our sample, 49 of which in MOS mode, for a total of more than 1000
galaxy redshifts.  This rate of data acquisition is in schedule with our
initial predictions, and by the beginning of 1995 we expect to be able
to complete a first `bright' sample of $\sim 400$ clusters with a flux 
cut of 0.15 cts s$^{-1}$.   We also expect the final sample to be
finished by 1996.  

During the observations, we serendipitously discovered two new gravitational
arc systems, shown in Fig.~2.  Although we routinely collect only
very short direct exposures (typically one minute), of those clusters which 
are candidates for MOS, these were enough for detecting the new 
arcs.   These results are a further indication of the potential advantage 
of searching for arcs in X--ray selected clusters, as also shown by other 
recent works (e.g. ref.~24).

\vspace{8cm}
\fcaption{{\tenit Left:} B image of the supplementary Abell cluster S295, with
exposure of 600~s, over a 2.6 arcmin square
field centered on the X--ray position\cite{edge}.  The gravitational arc, with a
redshift of 0.93 \cite{fort}, is the Mexican--hat--like feature next
to the elliptical galaxy northwest of the center. {\tenit Right:}
R image of the ROSAT cluster RXJ 1347.5-1145 over a 1.4 arcmin
square field\cite{sch}.  The two arclets are evident northeast and southwest 
of the central galaxy.}

\medskip
It is relevant to mention here that in the northern sky, a comparable redshift 
survey is being performed by Huchra and collaborators (see ref.~16).
It is planned that at some point both hemispheres will be 
homogenized to common selection criteria, and the two surveys
combined. This will provide an all--sky sample of $\sim 1500$ clusters with a
typical depth of $\sim 500$ $\hmpc$, which will be invaluable for
studies of large--scale structure, in particular the dipole of the 
cluster distribution.
\vspace{0.3cm}

We would like to thank the ROSAT team for producing such an exciting 
wealth of data, and ESO for providing us with the means to follow up with 
this enterprise.  This project would have never been possible outside the
``ESO Key--Programme" framework.

\end{document}